\documentclass[12pt]{article}
\usepackage{amssymb,amsmath,epsfig}
\allowdisplaybreaks

\begin{document}

\title{\bf Stability Analysis of Some Reconstructed Cosmological Models in $f(\mathcal{G},T)$ Gravity}
\author{M. Sharif \thanks {msharif.math@pu.edu.pk} and Ayesha Ikram
\thanks{ayeshamaths91@gmail.com}\\
Department of Mathematics, University of the Punjab,\\
Quaid-e-Azam Campus, Lahore-54590, Pakistan.}

\date{}

\maketitle

\begin{abstract}
The aim of this paper is to reconstruct and analyze the stability of
some cosmological models against linear perturbations in
$f(\mathcal{G},T)$ gravity ($\mathcal{G}$ and $T$ represent the
Gauss-Bonnet invariant and trace of the energy-momentum tensor,
respectively). We formulate the field equations for both general as
well as particular cases in the context of isotropic and homogeneous
universe model. We reproduce the cosmic evolution corresponding to
de Sitter universe, power-law solutions and phantom/non-phantom eras
in this theory using reconstruction technique. Finally, we study
stability analysis of de Sitter as well as power-law solutions
through linear perturbations.
\end{abstract}

\section{Introduction}

Modified theories of gravity have attained much attention after the
discovery of expanding accelerated universe. The basic ingredient
responsible for this tremendous change in cosmic history is some
mysterious type force having repulsive nature dubbed as dark energy.
The enigmatic nature of this energy has motivated many researchers
to unveil its hidden characteristics which are still not known.
Modified gravity approach is considered as the promising and
optimistic scenario among several other proposals that have been
presented to explore the salient features of dark energy. These
modified theories are established by adding or replacing curvature
invariants and their corresponding generic functions in the
Einstein-Hilbert action.

Lovelock theory of gravity is the direct generalization of general
relativity (GR) in $n$-dimensions which coincides with GR in
$4$-dimensions \cite{1}. The Ricci scalar $(R)$ is known as first
Lovelock scalar while Gauss-Bonnet (GB) invariant is the second
Lovelock scalar yielding Einstein-Gauss-Bonnet gravity in
$5$-dimensions \cite{2}. The GB invariant is a linear combination
with an interesting feature that it is free from spin-2 ghost
instabilities defined as \cite{3}
\begin{equation}\nonumber
\mathcal{G}=R^2-4R^{\alpha\beta}R_{\alpha\beta}+R^{\alpha\beta\mu\nu}
R_{\alpha\beta\mu\nu},
\end{equation}
where $R_{\alpha\beta}$ and $R_{\alpha\beta\mu\nu}$ are the Ricci
and Riemann tensors, respectively. This quadratic curvature
invariant is a topological term in 4-dimensions which possesses
trivial contribution in the field equations. To discuss the dynamics
of GB invariant in 4-dimensions, there are two interesting scenarios
either to couple $\mathcal{G}$ with scalar field or to add generic
function $f(\mathcal{G})$ in the Einstein-Hilbert action. The first
scheme naturally appears in the effective action in string theory
which investigates singularity-free cosmological solutions \cite{4}.
The second approach known as $f(\mathcal{G})$ gravity is introduced
as an alternative for dark energy which successfully discusses the
late-time cosmological evolution \cite{5}. This modified theory of
gravity is endowed with a quite rich cosmological structure as well
as consistent with solar system constraints \cite{6}.

The current cosmic accelerated expansion has also been discussed in
modified theories of gravity involving the curvature-matter
coupling. Harko et al. \cite{7} established $f(R,T)$ gravity to
study the curvature-matter coupling. Recently, we introduced the
curvature-matter coupling in $f(\mathcal{G})$ gravity named as
$f(\mathcal{G},T)$ theory of gravity \cite{8}. This coupling yields
non-zero covariant divergence of the energy-momentum tensor and an
extra force appears due to which massive test particles follow
non-geodesic trajectories while geodesic lines of geometry are
followed by the dust particles. Shamir and Ahmad \cite{9}
constructed some cosmologically viable models in $f(\mathcal{G},T)$
gravity using Noether symmetry approach. It is mentioned here that
cosmic expansion can be obtained from geometric as well as matter
components in such coupling.

The reconstruction as well as stability of cosmic evolutionary
models in modified theories of gravity are the captivating issues in
cosmology. In reconstruction technique, any known cosmic solution is
used in the modified field equations to find the corresponding
function which reproduces the given evolutionary cosmic history. In
stability analysis, the isotropic and homogeneous perturbations are
usually considered in which Hubble parameter as well as energy
density are perturbed to examine the background stability as time
evolves \cite{10}. Nojiri et al. \cite{11} formulated the
reconstruction scheme to reproduce some cosmological models in
$f(R)$ gravity. Elizalde et al. \cite{12} applied the same scenario
for $\Lambda$CDM cosmology ($\Lambda$ denotes cosmological constant
while CDM stands for cold dark matter) in $f(R,\mathcal{G})$ gravity
as well as in modified GB theories of gravity. The stability of
power-law solutions are also discussed in modified gravity theories
\cite{13}.

S\'{a}ez-G\'{o}mez \cite{14} explored the cosmological solutions in
$f(R)$ Ho$\check{\mathrm{r}}$ava-Lifshitz gravity and analyzed their
stability against first order perturbations around FRW universe.
Myrzakulov and his collaborators \cite{15} discussed the
cosmological models and found that $f(\mathcal{G})$ gravity could
successfully explain the cosmic evolutionary history. Jamil et al.
\cite{16} reconstructed the cosmological models in $f(R,T)$ gravity
and found that numerical analysis for Hubble parameter is in good
agreement with observational data for redshift parameter $<2$. The
stability of de Sitter, power-law solutions as well as $\Lambda$CDM
are analyzed in the context of $f(R,\mathcal{G})$ gravity \cite{17}.
Salako et al. \cite{18} studied the cosmological reconstruction,
stability as well as thermodynamics including first and second laws
for $\Lambda$CDM model in generalized teleparallel theory of
gravity. Sharif and Zubair \cite{19} demonstrated that $f(R,T)$
gravity can reproduce $\Lambda$CDM model, phantom or non-phantom
eras, de Sitter universe and power-law cosmic history. They also
analyzed the stability of reconstructed de Sitter as well as
power-law solutions.

In this paper, we reconstruct various cosmological models including
de Sitter universe, power-law solutions and phantom/non-phantom eras
in $f(\mathcal{G},T)$ theory. We also analyze the stability against
linear homogeneous perturbations for de Sitter as well as power-law
solutions. The paper has the following format. In section
\textbf{2}, we formulate the modified field equations while section
\textbf{3} is devoted to reconstruct some known cosmological
solutions in this gravity. Section \textbf{4} analyzes the stability
of specific solutions against linear perturbations around FRW
universe model. The results are summarized in the last section.

\section{$f(\mathcal{G},T)$ Gravity}

The action for $f(\mathcal{G},T)$ gravity is defined as \cite{8}
\begin{equation}\label{1}
\mathcal{I}=\int\left(\frac{R+f(\mathcal{G},T)}{2\kappa^2}
+\mathcal{L}_{m}\right)\sqrt{-g}d^{4}x,
\end{equation}
where $\kappa^2,~g$ and $\mathcal{L}_{m}$ represent coupling
constant, determinant of the metric tensor ($g_{\alpha\beta}$) and
Lagrangian associated with matter distribution, respectively.
Varying Eq.(\ref{1}) with respect to $g_{\alpha\beta}$, we obtain
the field equations
\begin{eqnarray}\nonumber
&&\kappa^2T_{\alpha\beta}-R_{\alpha\beta}+\frac{1}{2}g_{\alpha\beta}R
+\frac{1}{2}g_{\alpha\beta}f(\mathcal{G},T)-(T_{\alpha\beta}+
\Theta_{\alpha\beta})f_{T}(\mathcal{G},T)\\\nonumber&-&
[2RR_{\alpha\beta}-4R^{\mu}_{\alpha}R_{\mu\beta}
-4R_{\alpha\mu\beta\nu}R^{\mu\nu}+2R_{\alpha}^{\mu\nu\xi}
R_{\beta\mu\nu\xi}]f_{\mathcal{G}}(\mathcal{G},T)
\\\nonumber&-&[2Rg_{\alpha\beta}
\Box-4R_{\alpha\beta}\Box-2R\nabla_{\alpha}\nabla_{\beta}
+4R^{\mu}_{\beta}\nabla_{\alpha}\nabla_{\mu}
+4R^{\mu}_{\alpha}\nabla_{\beta}\nabla_{\mu}\\\label{2}&-&
4g_{\alpha\beta}R^{\mu\nu}
\nabla_{\mu}\nabla_{\nu}+4R_{\alpha\mu\beta\nu}
\nabla^{\mu}\nabla^{\nu}]f_{\mathcal{G}}(\mathcal{G},T)=0,
\end{eqnarray}
where $f_{T}(\mathcal{G},T)=\partial f(\mathcal{G},T)/\partial
T,~f_{\mathcal{G}}(\mathcal{G},T)=\partial
f(\mathcal{G},T)/\partial\mathcal{G},~\Box=\nabla_{\alpha}\nabla^{\alpha}$
($\nabla_{\alpha}$ denotes a covariant derivative) and
$T_{\alpha\beta}$ is the energy-momentum tensor. The expressions for
$T_{\alpha\beta}$ and $\Theta_{\alpha\beta}$ are \cite{20}
\begin{equation}\nonumber
T_{\alpha\beta}=g_{\alpha\beta}\mathcal{L}_{m}-2\frac{\partial
\mathcal{L}_{m}}{\partial g^{\alpha\beta}},\quad\Theta_{\alpha\beta}
=-2T_{\alpha\beta}+g_{\alpha\beta}\mathcal{L}_{m}-2g^{\mu\nu}
\frac{\partial^{2}\mathcal{L}_{m}}{\partial g^{\alpha\beta}\partial
g^{\mu\nu}},
\end{equation}
where we have assumed that $\mathcal{L}_{m}$ depends only on
$g_{\alpha\beta}$ rather than its derivatives. The non-zero
divergence of $T_{\alpha\beta}$ is given by
\begin{eqnarray}\nonumber
\nabla^{\alpha}T_{\alpha\beta}&=&\frac{1}{\kappa^2-f_{T}(\mathcal{G},T)}
\left[\left(\nabla^{\alpha}\Theta_{\alpha\beta}
-\frac{1}{2}g_{\alpha\beta}\nabla^{\alpha}T\right)f_{T}(\mathcal{G},T)
+(\Theta_{\alpha\beta}\right.\\\label{4}&+&\left.
T_{\alpha\beta})\nabla^{\alpha}f_{T}(\mathcal{G},T)\right].
\end{eqnarray}
The above equations indicate that the complete dynamics of
$f(\mathcal{G},T)$ gravity is based on the suitable choice of
$\mathcal{L}_{m}$.

The energy-momentum tensor for perfect fluid is
\begin{equation}\label{5}
T_{\alpha\beta}=(\rho+p)u_{\alpha}u_{\beta}-pg_{\alpha\beta},
\end{equation}
where $u_{\alpha},\rho$ and $p$ represent the four velocity, energy
density and pressure of matter distribution, respectively. In this
case, the expression for $\Theta_{\alpha\beta}$ becomes
\begin{equation}\label{6}
\Theta_{\alpha\beta}=-pg_{\alpha\beta}-2T_{\alpha\beta},
\end{equation}
where $\mathcal{L}_{m}=-p$. The line element for FRW universe model
is given by
\begin{equation}\label{7}
ds^2=dt^2-a^2(t)(dx^2+dy^2+dz^2),
\end{equation}
where $a(t)$ is the scale factor. Using Eqs.(\ref{5})-(\ref{7}) in
(\ref{2}), we obtain the corresponding field equation as follows
\begin{eqnarray}\nonumber
3H^2&=&\kappa^2\rho+\frac{1}{2}f(\mathcal{G},T)+(\rho+p)
f_{T}(\mathcal{G},T)-12H^2(H^2+\dot{H})f_{\mathcal{G}}(\mathcal{G},T)
\\\label{8}&+&12H^3\dot{f}_{\mathcal{G}}(\mathcal{G},T),
\end{eqnarray}
where
$H=\frac{\dot{a}}{a},~T=\rho-3p,~\mathcal{G}=24H^2(\dot{H}+H^2)$ and
dot represents derivative with respect to time. The non-zero
continuity equation (\ref{4}) takes the form
\begin{equation}\label{9}
\dot{\rho}+3H(\rho+p)=\frac{-1}{\kappa^2+f_{T}(\mathcal{G},T)}\left[
\left(\dot{p}+\frac{1}{2}\dot{T}\right)f_{T}(\mathcal{G},T)+
(\rho+p)\dot{f}_{T}(\mathcal{G},T)\right].
\end{equation}
The standard conservation law holds if right hand side of this
equation vanishes. For equation of state $p=\omega\rho$ ($\omega$ is
the equation of state parameter), Eq.(\ref{9}) yields
\begin{equation}\label{10}
\dot{\rho}=-3H(1+\omega)\rho,
\end{equation}
with additional constraint
\begin{eqnarray}\label{11}
\frac{1}{2}\dot{\rho}(1-\omega)f_{T}+\rho(1+\omega)\left(
\dot{\mathcal{G}}f_{\mathcal{G}T}+\dot{T}f_{TT}\right)=0.
\end{eqnarray}

We rewrite the above equations in terms of new variable
$\mathcal{N}$ known as e-folding instead of $t$ which is also
related with redshift parameter $(z)$ as \cite{11}
\begin{equation}\nonumber
\mathcal{N}=-\ln(1+z)=\ln{(a/a_{0})}.
\end{equation}
Using the above definition of $\mathcal{N}$, Eqs.(\ref{8}) and
(\ref{9}) become
\begin{eqnarray}\nonumber
3H^2&=&\kappa^2\rho+\frac{1}{2}f+\rho(1+\omega)f_{T}-12H^3(H+H')f_{\mathcal{G}}
+288H^6\\\label{13}&\times&(HH''+3H'^{2}+4HH')f_{\mathcal{GG}}+12H^4T'
f_{\mathcal{G}T},\\\nonumber\rho'+3(1+\omega)\rho&=&\frac{-1}{\kappa^2+f_{T}}
\left[\left(\omega\rho'+\frac{1}{2}T'\right)f_{T}+\rho(1+\omega)
\left(\mathcal{G}'f_{\mathcal{G}T}+T'f_{TT}\right)\right],
\end{eqnarray}
where $H=d\mathcal{N}/dt,~d/dt=H(d/d\mathcal{N})$ and prime denotes
derivative with respect to $\mathcal{N}$. The simplest choice of
$f(\mathcal{G},T)$ model is
\begin{equation}\label{15}
f(\mathcal{G},T)=F(\mathcal{G})+\mathcal{F}(T),
\end{equation}
which possesses no direct non-minimally coupling between curvature
and matter. For this particular model, the field equation (\ref{13})
splits into a set of two ordinary differential equations as
\begin{equation}\nonumber
288H^{6}(HH''+3H'^{2}+4HH')F_{\mathcal{GG}}-12H^{3}(H+H')F_{\mathcal{G}}
+\frac{1}{2}F(\mathcal{G})-3H^{2}=0,
\end{equation}
\begin{equation}\\\nonumber
\rho(1+\omega)\mathcal{F}_{T}+\frac{1}{2}\mathcal{F}(T)+\kappa^2\rho=0,
\end{equation}
where $F_{\mathcal{G}}=dF(\mathcal{G})/d\mathcal{G}$ and
$\mathcal{F}_{T}=d\mathcal{F}(T)/dT$. The field equations for
perfect fluid matter distribution in $f(\mathcal{G})$ gravity is
recovered if $\mathcal{F}(T)$ vanishes while GR is achieved for
$f(\mathcal{G},T)=0$.

\section{Cosmological Reconstruction}

In this section, we reproduce different cosmological scenarios
including de Sitter universe, power-law solutions and
phantom/non-phantom eras in $f(\mathcal{G},T)$ gravity.

\subsection{de Sitter Universe}

The de Sitter cosmic evolution is interesting and well-known as it
elegantly describes current expansion of the universe. This solution
is considered as the universe in which the energy density of matter
and radiation is negligible as compared to vacuum energy (energy
density for DE dominated era) and thus the universe expands forever
at a constant rate. The scale factor of this evolutionary model
grows exponentially with constant Hubble parameter $H(t)=H_{0}$,
defined as \cite{17}
\begin{equation}\label{1d}
a(t)=a_{0}e^{H_{0}t},
\end{equation}
where $a_{0}$ is an integration constant. Equation (\ref{10}) gives
energy density of the form
\begin{equation}\label{2d}
\rho=\rho_{0}e^{-3(1+\omega)H_{0}t},
\end{equation}
where $\omega\neq-1$ and $\rho_{0}$ is a constant. Using
Eqs.(\ref{1d}) and (\ref{2d}) in (\ref{8}), we obtain
\begin{eqnarray}\nonumber
&&\frac{1}{2}f(\mathcal{G}_{0},T)-12H_{0}^{4}f_{\mathcal{G}}(\mathcal{G}
_{0},T)+\left(\frac{1+\omega}{1-3\omega}\right)Tf_{T}(\mathcal{G}_{0},T)
-36(1+\omega)H_{0}^{4}T\\\label{3d}&\times&f_{\mathcal{G}T}(\mathcal{G}_{0},T)
+\frac{\kappa^2T}{1-3\omega}-3H_{0}^{2}=0,
\end{eqnarray}
where $\mathcal{G}_{0}=24H_{0}^4$ is the GB invariant at
$H(t)=H_{0}$. The solution of the above differential equation is
\begin{eqnarray}\nonumber
f(\mathcal{G},T)&=&c_{1}c_{2}e^{c_{1}\mathcal{G}}T^{-\frac{1}{2}
\left(\frac{(1-24c_{1}H_{0}^{4})(1-3\omega)}{1+\omega-36c_{1}
H_{0}^{4}(1-3\omega)}\right)}+c_{1}c_{2}T^{-\frac{1}{2}}\left(
\frac{1-3\omega}{1+\omega}\right)
\\\label{4d}&-&\frac{2\kappa^2}{3-\omega}T+6H_{0}^{2},
\end{eqnarray}
where $c_{i}$'s $(i=1,2)$ are integration constants. Since we have
used the continuity equation (\ref{10}) in Eq.(\ref{3d}), so we must
constrain its solution. Using the above equation with Eq.(\ref{11}),
we obtain the following functions
\begin{eqnarray}\label{5d}
f_{1}(\mathcal{G},T)&=&c_{1}c_{2}\Xi_{1}e^{c_{1}\mathcal{G}}T^{-\frac{1}{2}
\left(\frac{(1-24c_{1}H_{0}^{4})(1-3\omega)}{1+\omega-36c_{1}
H_{0}^{4}(1-3\omega)}\right)}+\frac{2\kappa^2\omega}{1-3\omega}T+6H_{0}^{2},
\\\label{6d} f_{2}(\mathcal{G},T)&=&c_{1}c_{2}\Xi_{2}T^{-\frac{1}{2}}\left(
\frac{1-3\omega}{1+\omega}\right)+\frac{2\kappa^2}{3-\omega}\Xi_{3}T+6H_{0}^{2},
\end{eqnarray}
where $\Xi_{j}$'s $(j=1,2,3)$ are constants in terms of $\omega$ and
$H_{0}$ given in Appendix \textbf{A}. For the model (\ref{15}), we
have
\begin{eqnarray}\label{7d}
3H_{0}^{2}-\frac{1}{2}F+12H_{0}^{4}F_{\mathcal{G}}=0,
\quad\kappa^2\rho+\frac{1}{2}\mathcal{F}+(1+\omega)\rho
\mathcal{F}_{T}=0,
\end{eqnarray}
where the first equation corresponds to de Sitter universe in the
absence of matter contents in $f(\mathcal{G})$ gravity \cite{6}.
Using the constraint (\ref{11}), the second equation becomes
\begin{equation}\label{8d}
\kappa^2(1-\omega)T+\frac{1}{2}(1-3\omega)(1-\omega)\mathcal{F}-2(1+\omega)^{2}
T^{2}\mathcal{F}_{TT}=0.
\end{equation}
The solution of Eqs.(\ref{7d}) and (\ref{8d}) leads to
\begin{eqnarray}\nonumber
f(\mathcal{G},T)&=&\hat{c}_{1}e^{\frac{\mathcal{G}}{24H_{0}^{4}}}+\hat{c}_{2}
T^{\frac{1}{2}\left(1+\frac{\sqrt{2(1-\omega+2\omega^2)}}{1+\omega}\right)}
+\hat{c}_{3}T^{\frac{1}{2}\left(1-\frac{\sqrt{2(1-\omega+2\omega^2)}}
{1+\omega}\right)}\\\label{9d}&-&\frac{2\kappa^2T}{1-3\omega}+6H_{0}^{2},
\end{eqnarray}
where $\hat{c}_{j}$'s are constants of integration. Equations
(\ref{4d}) and (\ref{9d}) indicate that de Sitter expansion can also
be described in $f(\mathcal{G},T)$ gravity.

\subsection{Power-law Solutions}

Power-law solutions have significant importance to discuss different
evolutionary phases of the universe in modified theory. These
solutions describe the decelerated as well as accelerated cosmic
eras which are characterized by the scale factor as \cite{17}
\begin{equation}\label{1p}
a(t)=a_{0}t^{\lambda},\quad H=\frac{\lambda}{t},\quad\lambda>0.
\end{equation}
The cosmic decelerated phase is observed for $0<\lambda<1$ including
the radiation $(\lambda=\frac{1}{2})$ as well as dust
$(\lambda=\frac{2}{3})$ dominated eras while $\lambda>1$ covers the
accelerated phase of the universe. For this scale factor, the GB
invariant takes the form
\begin{equation}\label{2p}
\mathcal{G}=24\frac{\lambda^3}{t^4}(\lambda-1).
\end{equation}
Using Eqs.(\ref{10}), (\ref{1p}) and (\ref{2p}), the field equation
becomes
\begin{eqnarray}\nonumber
&&\frac{1}{2}f-\frac{1}{2}\mathcal{G}f_{\mathcal{G}}+\frac{(1+\omega)T}
{1-3\omega}f_{T}-\frac{2}{\lambda-1}\mathcal{G}^2f_{\mathcal{GG}}-
\frac{2\lambda(1+\omega)\mathcal{G}T}{2(\lambda-1)}f_{\mathcal{G}T}
\\\label{3p}&-&3\lambda^{2}\left(\frac{T}{\rho_{0}(1-3\omega)}
\right)^{\frac{2}{3\lambda(1+\omega)}}+\frac{\kappa^2T}{1-3\omega}=0,
\end{eqnarray}
whose solution is given by
\begin{eqnarray}\label{4p}
f(\mathcal{G},T)=\tilde{c}_{1}\tilde{c}_{3}T^{\tilde{c}_{2}}
\mathcal{G}^{\frac{1}{4}(\gamma_{1}+\gamma_{2})}+
\tilde{c}_{2}\tilde{c}_{3}T^{\tilde{c}_{2}}\mathcal{G}^{\frac{1}{4}
(\gamma_{1}-\gamma_{2})}+\tilde{c}_{1}\tilde{c}_{2}
T^{\gamma_{3}}+\gamma_{4}T+\gamma_{5}T^{\gamma_{6}},
\end{eqnarray}
where $\tilde{c}_{j}$'s are integration constants and
$\gamma_{\hat{j}}$'s $(\hat{j}=1...6)$ are given in Appendix
\textbf{A}. Inserting Eq.(\ref{4p}) in (\ref{11}), we obtain
\begin{eqnarray}\label{5p}
f_{1}(\mathcal{G},T)&=&\tilde{c}_{1}\tilde{c}_{3}\Delta_{1}T^{\tilde{c}_{2}}
\mathcal{G}^{\frac{1}{4}(\gamma_{1}+\gamma_{2})}+\tilde{c}_{1}\tilde{c}_{2}
\Delta_{2}T^{\gamma_{3}}+\Delta_{3}T+\Delta_{4}T^{\gamma_{6}},\\\label{6p}
f_{2}(\mathcal{G},T)&=&\tilde{c}_{2}\tilde{c}_{3}\Omega_{1}
T^{\tilde{c}_{2}}\mathcal{G}^{\frac{1}{4}(\gamma_{1}-\gamma_{2})}
+\tilde{c}_{1}\tilde{c}_{2}\Omega_{2}T^{\gamma_{3}}+\Omega_{3}T
+\Omega_{4}T^{\gamma_{6}}.
\end{eqnarray}
where $\Delta_{k}$'s and $\Omega_{k}$'s $(k=1...4)$ are given in
Appendix \textbf{A}.

Now we find the expression of $f(\mathcal{G},T)$ for the choice of
model (\ref{15}). The differential equation (\ref{3p}) yields two
ordinary differential equations in variables $\mathcal{G}$ and $T$
given by
\begin{eqnarray}\nonumber
&&F-\mathcal{G}F_{\mathcal{G}}-\frac{4}{\lambda-1}\mathcal{G}^{2}
F_{\mathcal{GG}}=0,\\\nonumber&&\mathcal{F}-\frac{4(1+\omega)^{2}}{(1-\omega)(1-3\omega)}T^{2}
\mathcal{F}_{TT}+\frac{2\kappa^{2}T}{1-3\omega}-6\lambda^{2}
\left(\frac{T}{\rho_{0}(1-3\omega)}\right)^{\frac{2}{3\lambda
(1+\omega)}}=0.
\end{eqnarray}
The solution of these equations provide $f(\mathcal{G},T)$ model as
\begin{eqnarray}\nonumber
f(\mathcal{G},T)&=&\bar{c}_{1}\mathcal{G}+\bar{c}_{2}\mathcal{G}
^{\frac{1-\lambda}{4}}+\bar{c}_{3}T^{\frac{1}{2}\left(1+
\frac{\sqrt{2(1-\omega+2\omega^{2})}}{1+\omega}\right)}
+\bar{c}_{4}T^{\frac{1}{2}\left(1-\frac{\sqrt{2(1-\omega+2
\omega^{2})}}{1+\omega}\right)}\\\nonumber&-&\frac{2\kappa^{2}T}
{1-3\omega}+\frac{54\lambda^{4}(1-\omega)(1-3\omega)}{9\lambda^{2}
(1-\omega)(1-3\omega)-8[2-3\lambda(1+\omega)]}\\\label{7p}&\times&
\left(\frac{T}{\rho_{0}(1-3\omega)}\right)^{\frac{2}{3\lambda(1+\omega)}},
\end{eqnarray}
where $\bar{c}_{k}$'s are integration constants. Thus, the power-law
solutions are reconstructed which may be helpful to explore the
expansion history of the universe in this modified theory of
gravity.

\subsection{Phantom and non-Phantom Matter Fluids}

Here, we reconstruct $f(\mathcal{G},T)$ model which can explain the
system including both phantom and non-phantom eras. In the Einstein
gravity, the Hubble parameter describing the phantom as well as
non-phantom matter distribution is given by \cite{11}
\begin{equation}\label{1q}
H^2=\frac{\kappa^2}{3}(\rho_{p}a^{b}+\rho_{q}a^{-b}),
\end{equation}
where $b,~\rho_{p}$ and $\rho_{q}$ represent the model parameter,
energy densities of phantom and non-phantom matter fluids,
respectively. The violation of all four energy conditions leads to
phantom phase and the energy density grows while it decreases in a
non-phantom regime. The phantom energy density would become infinite
in finite time, causing a huge gravitational repulsion that the
universe would lose all structure and end in a big-rip singularity
\cite{aa}. When the scale factor is large, the first term on right
hand side dominates which corresponds to the phantom era of the
universe with $\omega=-1-b/3<-1$. The non-phantom era in the early
universe is observed for $\omega=-1+b/3>-1$ when the scale factor is
small and the second term dominates. We rewrite $H(t)$ in terms of a
new function $S(\mathcal{N})$ as $H^{2}=S(\mathcal{N})$ so that
Eq.(\ref{1q}) becomes
\begin{equation}\label{2q}
S(\mathcal{N})=S_{p}e^{b\mathcal{N}}+S_{q}e^{-b\mathcal{N}},
\end{equation}
where $S_{p}=\frac{\kappa^2}{3}\rho_{p}a_{0}^{b}$ and
$S_{q}=\frac{\kappa^2}{3}\rho_{q}a_{0}^{-b}$. The GB invariant takes
the form
\begin{equation}\label{3q}
\mathcal{G}=24S^{2}(\mathcal{N})+12S(\mathcal{N})S'(\mathcal{N}).
\end{equation}
Inserting Eq.(\ref{2q}) in (\ref{3q}), we obtain a quadratic
equation in $e^{2b\mathcal{N}}$ whose solution is given by
\begin{equation}\nonumber
e^{2b\mathcal{N}}=\frac{-(48S_{p}S_{q}-\mathcal{G})\pm\sqrt{(48S_{p}
S_{q}-\mathcal{G})^{2}-576(4-b^2)S_{p}^{2}S_{q}^{2}}}{24(2+b)
S_{p}^{2}},\quad b\neq2.
\end{equation}
For the sake of simplicity, we consider $b=2$ so that it reduces to
\begin{equation}\label{4q}
e^{2b\mathcal{N}}=\frac{\mathcal{G}-48S_{p}S_{q}}{48S_{p}^{2}}.
\end{equation}
Using Eqs.(\ref{2q}) and (\ref{4q}) in (\ref{8}), we have
\begin{eqnarray}\nonumber
&&\frac{1}{2}f-\frac{1}{2}\mathcal{G}f_{\mathcal{G}}+\left(\frac{1+\omega}
{1-3\omega}\right)Tf_{T}+\mathcal{G}^{2}f_{\mathcal{GG}}-\frac{3(1+\omega)
\mathcal{G}^{2}T}{4(\mathcal{G}-48S_{p}S_{q})}f_{\mathcal{G}T}\\\nonumber&-&
\frac{1}{4}\sqrt{\frac{3\mathcal{G}^{2}}{\mathcal{G}-48S_{p}S_{q}}}
+\frac{\kappa^2T}{1-3\omega}=0,
\end{eqnarray}
which is a complicated partial differential equation whose
analytical solution cannot be found.

To find the reconstructed $f(\mathcal{G},T)$ model, we consider its
particular form (\ref{15}) which provides the following set of
differential equations
\begin{eqnarray}\nonumber
&&\frac{1}{2}F-\frac{1}{2}\mathcal{G}F_{\mathcal{G}}
+\mathcal{G}^{2}F_{\mathcal{GG}}-\frac{1}{4}\sqrt{\frac{3\mathcal{G}^{2}}
{\mathcal{G}-48S_{p}S_{q}}}=0,\\\nonumber&&\mathcal{F}-\frac{4
(1+\omega)^{2}}{(1-\omega)(1-3\omega)}T^{2}\mathcal{F}_{TT}
+\frac{2\kappa^{2}T}{1-3\omega}=0,
\end{eqnarray}
where we have used the additional constraint in the second equation.
Solving these equations, it follows that
\begin{eqnarray}\nonumber
f(\mathcal{G},T)&=&d_{1}\mathcal{G}^{\frac{1}{2}}+d_{2}\mathcal{G}
+d_{3}T^{\frac{1}{2}\left(1+\frac{\sqrt{2(1-\omega+2\omega^{2})}}
{1+\omega}\right)}+d_{4}T^{\frac{1}{2}\left(1-\frac{\sqrt{2(1-\omega
+2\omega^{2})}}{1+\omega}\right)}\\\nonumber&+&\frac{1}
{4\sqrt{S_{p}S_{q}}}\left[\mathcal{G}\tan^{-1}\left(\frac{1}{12}
\sqrt{\frac{3(\mathcal{G}-S_{p}S_{q})}{S_{p}S_{q}}}\right)-
2\sqrt{3S_{p}S_{q}\mathcal{G}}\right.\\\label{6q}&\times&\left.
\ln{\left(\mathcal{G}-24S_{p}S_{q}+\sqrt{\mathcal{G}
(\mathcal{G}-48S_{p}S_{q})}\right)}\right]-\frac{2\kappa^{2}T}{1-3\omega}.
\end{eqnarray}
where $d_{k}$'s are constants of integration. Thus, phantom and
non-phantom cosmic history can be discussed in $f(\mathcal{G},T)$
gravity.

\section{Perturbations and Stability of Cosmological Solutions}

In this section, we analyze stability of some cosmological
evolutionary solutions about linear homogeneous perturbations in
this modified gravity. We construct the perturbed field as well as
continuity equations using isotropic and homogeneous universe model
for both general and particular cases including de Sitter and
power-law solutions. We assume a general solution
\begin{equation}\label{1t}
H(t)=H_{*}(t),
\end{equation}
which satisfies the basic field equations for FRW universe model in
$f(\mathcal{G},T)$ gravity. In terms of the above solution, the
expressions for $\mathcal{G}_{*}$ and $T_{*}$ are
\begin{eqnarray}\nonumber
\mathcal{G}_{*}=24H_{*}^{2}(H_{*}^{2}+\dot{H}_{*})=24H_{*}^{3}(H_{*}
+H'_{*}),\quad T_{*}=\rho_{*}(t)(1-3\omega).
\end{eqnarray}
For any particular $f(\mathcal{G},T)$ model that can regenerate the
above solution (\ref{1t}), the following equation of motion as well
as non-zero divergence of the energy-momentum tensor must be
satisfied
\begin{eqnarray}\nonumber
3H_{*}^{2}&=&\kappa^2\rho_{*}+(1+\omega)\rho_{*}f_{T}^{*}+\frac{1}{2}
f^{*}-12H_{*}^{3}(H_{*}+H'_{*})f_{\mathcal{G}}^{*}+288\\\nonumber
&\times&H_{*}^{6}(H_{*}H''_{*}+3H_{*}^{'2}+4H_{*}H'_{*})
f_{\mathcal{GG}}^{*}+12H_{*}^{4}T'_{*}f_{\mathcal{G}T}^{*},\\\nonumber
\rho'_{*}+3(1+\omega)\rho_{*}&=&\frac{-1}{\kappa^2+f_{T}^{*}}\left[
\frac{1}{2}(T'_{*}+2\omega\rho'_{*})f_{T}^{*}+(1+\omega)\rho_{*}
\left(\mathcal{G}'_{*}f_{\mathcal{G}T}^{*}\right.
\right.\\\nonumber&+&\left.\left.T'_{*}f_{TT}^{*}\right)\right],
\end{eqnarray}
where superscript $*$ denotes that the function and its
corresponding derivatives are calculated at
$\mathcal{G}=\mathcal{G}_{*}$ and $T=T_{*}$. If the conservation law
holds, we get energy density in terms of $H_{*}(t)$ as
\begin{equation}\nonumber
\rho_{*}(t)=\rho_{0}e^{-3(1+\omega)\int H_{*}(t)dt}.
\end{equation}
The first order perturbations in Hubble parameter and energy density
are defined as
\begin{equation}\label{6t}
H(t)=H_{*}(t)(1+\delta(t)),\quad\rho(t)=\rho_{*}(t)(1+\delta_{m}(t)),
\end{equation}
where $\delta(t)$ and $\delta_{m}(t)$ are the perturbation
parameters.

In order to analyze first order perturbations about the solution
(\ref{1t}), we apply the series expansion on the function
$f(\mathcal{G},T)$ as
\begin{equation}\label{7t}
f(\mathcal{G},T)=f^{*}+f_{\mathcal{G}}^{*}(\mathcal{G}-\mathcal{G}_{*})
+f_{T}^{*}(T-T_{*})+\mathcal{O}^{2},
\end{equation}
where $\mathcal{O}^{2}$ involves the terms proportional to quadratic
or higher powers of $\mathcal{G}$ and $T$ while only the linear
terms are considered. Using Eqs.(\ref{6t}) and (\ref{7t}) in
(\ref{8}), we obtain the following perturbed field equation
\begin{equation}\label{8t}
\chi_{1}\ddot{\delta}+\chi_{2}\dot{\delta}+\chi_{3}\delta+
\chi_{4}\dot{\delta}_{m}+\chi_{5}\delta_{m}=0,
\end{equation}
where $\chi_{h}$'s $(h=1...5)$ are given in Appendix \textbf{A}.
Inserting these perturbations in Eq.(\ref{9}), the perturbed
continuity equation is
\begin{equation}\label{9t}
\Upsilon_{1}\delta+\Upsilon_{2}\dot{\delta}+\Upsilon_{3}\ddot{\delta}
+\Upsilon_{4}\delta_{m}+\Upsilon_{5}\dot{\delta}_{m}=0,
\end{equation}
where $\Upsilon_{h}$'s are provided in Appendix \textbf{A}. If the
conversation law holds in this modified gravity, Eq.(\ref{9t})
reduces to
\begin{equation}\label{10t}
\dot{\delta}_{m}+3(1+\omega)H_{*}\delta=0.
\end{equation}
The perturbed equations (\ref{8t}) and (\ref{9t}) are helpful to
analyze the stability of any specific FRW cosmological evolutionary
model in $f(\mathcal{G},T)$ gravity. For the particular model
(\ref{15}), these perturbed equations reduce to
\begin{eqnarray}\nonumber
\hat{\chi}_{1}\ddot{\delta}+\hat{\chi}_{2}\dot{\delta}+\hat{\chi}_{3}
\delta+\hat{\chi}_{5}\delta_{m}=0,\\\nonumber
\hat{\Upsilon}_{1}\delta+\hat{\Upsilon}_{4}\delta_{m}
+\hat{\Upsilon}_{5}\dot{\delta}_{m}=0,
\end{eqnarray}
where the coefficients of $(\delta,~\delta_{m})$ and their
derivatives are expressed in Appendix \textbf{A}. In the following
subsections, we investigate the stability of de Sitter and power-law
solutions.

\subsection{Stability of de Sitter Solutions}

Consider the de Sitter solution $H_{*}(t)=H_{0}$, the perturbed
equation (\ref{8t}) takes the form
\begin{eqnarray}\nonumber
&&288H_{0}^{6}f_{\mathcal{GG}}^{0}\ddot{\delta}+\left(864H_{0}^{7}
f_{\mathcal{GG}}^{0}+24\rho_{*}H_{0}^{3}(1+\omega)f_{\mathcal{G}T}^{0}
-864\rho_{*}H_{0}^{7}(1-3\omega)\right.\\\nonumber&\times&\left.
(1+\omega)f_{\mathcal{GG}T}^{0}\right)\dot{\delta}+\left(-6H_{0}^{2}
-1152H_{0}^{8}f_{\mathcal{GG}}^{0}+12\rho_{*}H_{0}^{2}(1+\omega)
[8H_{0}^{2}\right.\\\nonumber&-&\left.9H_{0}^{2}(1-3\omega)]
f_{\mathcal{G}T}^{0}-3456\rho_{*}H_{0}^{8}(1-3\omega)(1+\omega)
f_{\mathcal{GG}T}^{0}\right)\delta+12\rho_{*}H_{0}^{3}\\\nonumber
&\times&(1-3\omega)f_{\mathcal{G}T}^{0}\dot{\delta}_{m}+\left(\kappa^{2}
\rho_{*}+\frac{1}{2}\rho_{*}(3-\omega)f_{T}^{0}+\rho_{*}^{2}(1-3\omega)
(1+\omega)f_{TT}^{0}\right.\\\nonumber&-&\left.12\rho_{*}
H_{0}^{2}(1-3\omega)[H_{0}^{2}+3(1+\omega)H_{0}^{2}]f_{\mathcal{G}T}^{0}
-36\rho_{*}^{2}H_{0}^{4}(1-3\omega)^{2}(1+\omega)
\right.\\\label{13t}&\times&\left.f_{\mathcal{G}TT}^{0}\right)\delta_{m}=0,
\end{eqnarray}
where the superscript $0$ represents that the function and its
corresponding derivatives are evaluated at $\mathcal{G}_{0}$ and
$T_{0}$. We consider the conserved perturbed equation for stability
analysis since the de Sitter solutions are constructed using the
constraint (\ref{11}) in the previous section. The numerical
technique is used to solve Eqs.(\ref{10t}) and (\ref{13t}) for the
model (\ref{5d}). The evolution of $\delta(t)$ and $\delta_{m}(t)$
are shown in Figure \textbf{1}. We consider $H_{0}=67.8$ and
$\kappa=1$ throughout the stability analysis of de Sitter universe
models whereas integration constants are $c_{1}=1\times10^{-6}$ and
$c_{2}=-1\times10^{-3}$.
\begin{figure}
\epsfig{file=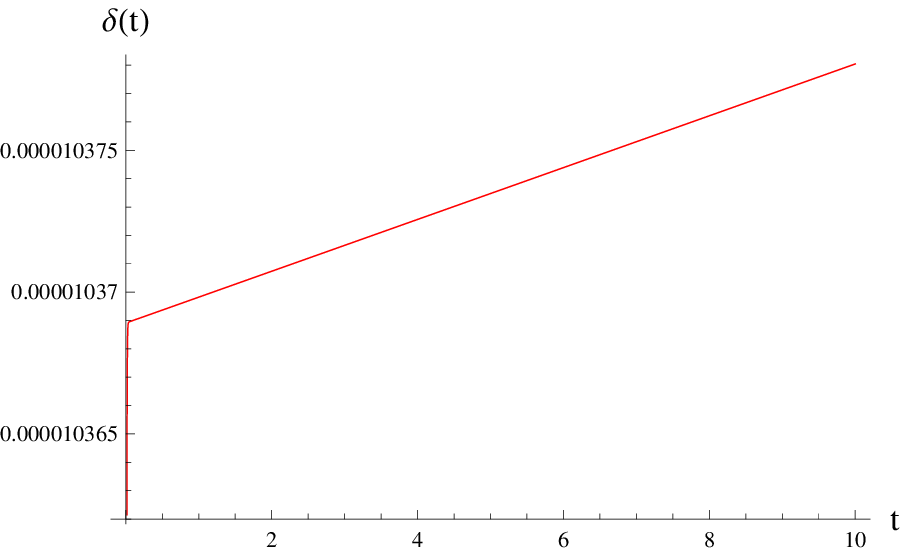, width=0.5\linewidth}\epsfig{file=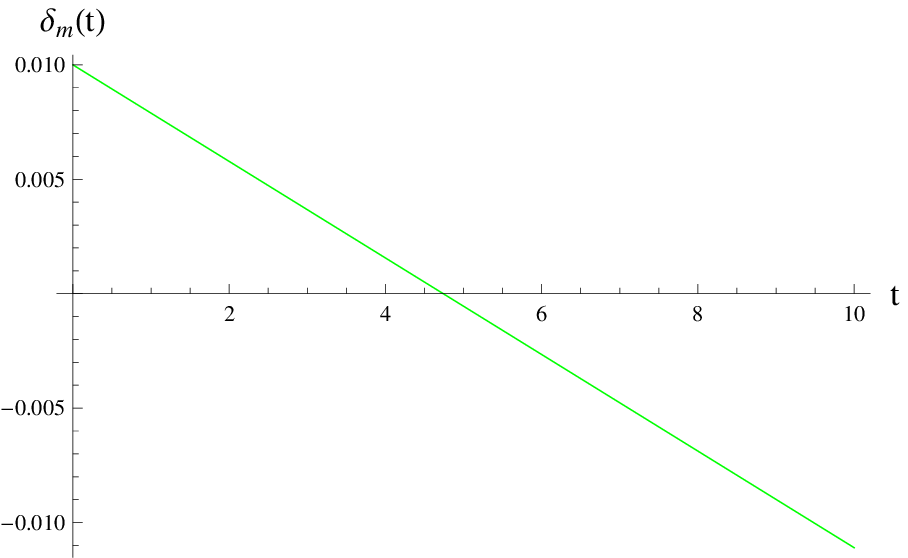,
width=0.5\linewidth}\caption{Evolution of perturbations $\delta(t)$
and $\delta_{m}(t)$ for model (\ref{5d}) with $\omega=0$.}
\end{figure}
\begin{figure}
\epsfig{file=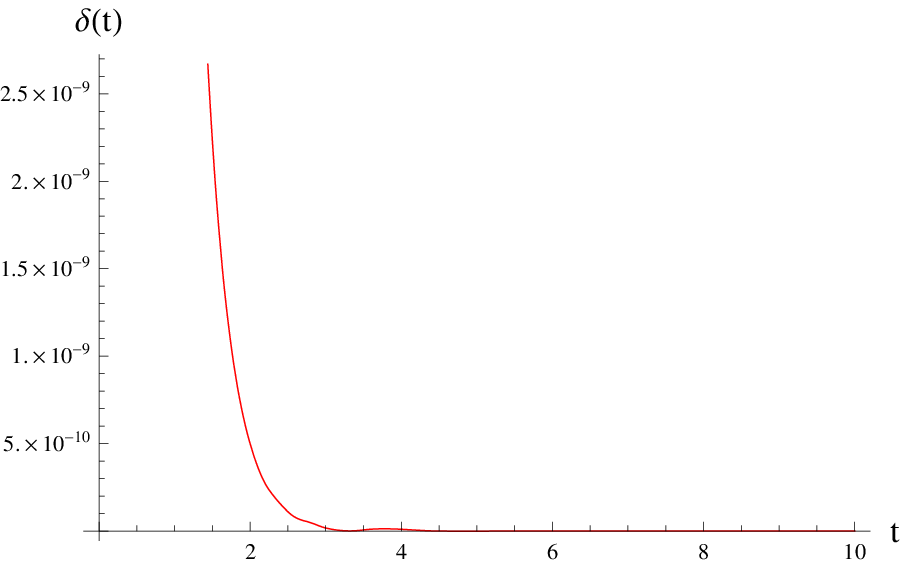, width=0.5\linewidth}\epsfig{file=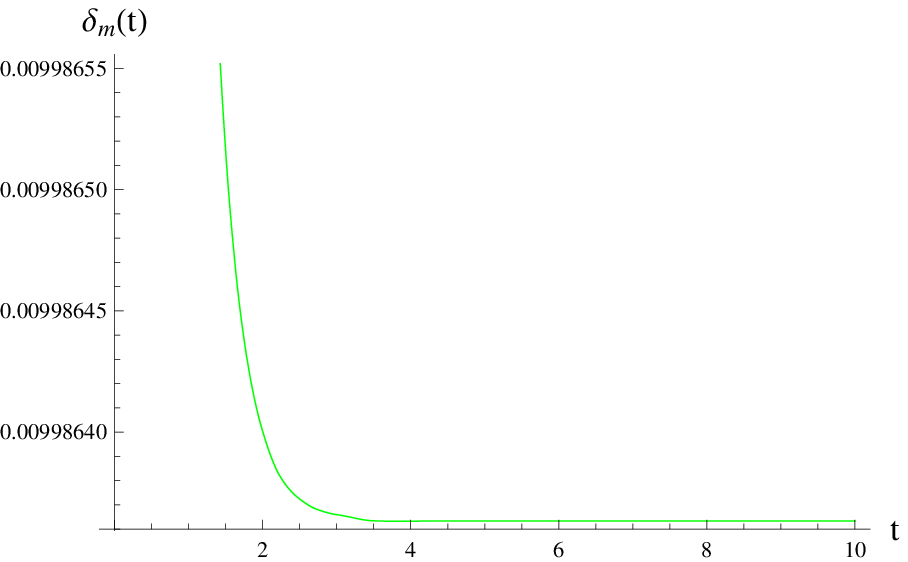,
width=0.5\linewidth}\caption{Evolution of perturbations $\delta(t)$
and $\delta_{m}(t)$ for model (\ref{6d}) with $\omega=0$.}
\end{figure}

Figure \textbf{1} shows smooth behavior of $\delta(t)$ (left) and
$\delta_{m}(t)$ (right) which do not decay in late times indicating
that de Sitter model (\ref{5d}) is unstable. The stability analysis
of model (\ref{6d}) with same integration constants is shown in
Figure \textbf{2}. In the left panel, it is observed that small
oscillations are produced about $t=4$ while it decays in late times,
thus the model (\ref{6d}) shows stable behavior against
perturbations. For model (\ref{15}), Eq.(\ref{13t}) becomes
\begin{eqnarray}\nonumber
&&288H_{0}^{6}F_{\mathcal{GG}}^{0}\ddot{\delta}+864H_{0}^{7}F_{\mathcal{GG}}
^{0}\dot{\delta}
+\left(-6H_{0}^{2}-1152H_{0}^{8}F_{\mathcal{GG}}^{0}\right)\delta
\\\label{14t}&+&\left(\kappa^{2}\rho_{*}+\frac{1}{2}\rho_{*}(3-\omega)
\mathcal{F}_{T}^{0}+\rho_{*}^{2}(1-3\omega)(1+\omega)
\mathcal{F}_{TT}^{0}\right)\delta_{m}=0.
\end{eqnarray}
Figure \textbf{3} represents the behavior of $\delta(t)$ and
$\delta_{m}(t)$ for model (\ref{9d}) with integration constants
$\hat{c}_{1}=-10,~\hat{c}_{2}=0.001$ and $\hat{c}_{3}=1$. It is
shown that oscillations in perturbation parameters are produced
initially as shown in Figure \textbf{3}. This oscillating behavior
is clearly observed in Figure \textbf{4} which decays in future for
both $\delta(t)$ as well as $\delta_{m}(t)$ and hence the solution
becomes stable.
\begin{figure}
\epsfig{file=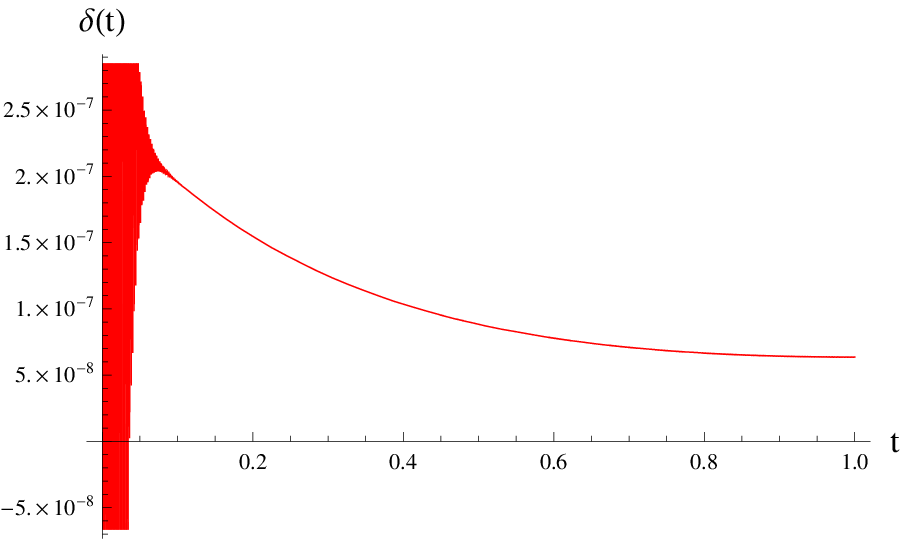, width=0.5\linewidth}\epsfig{file=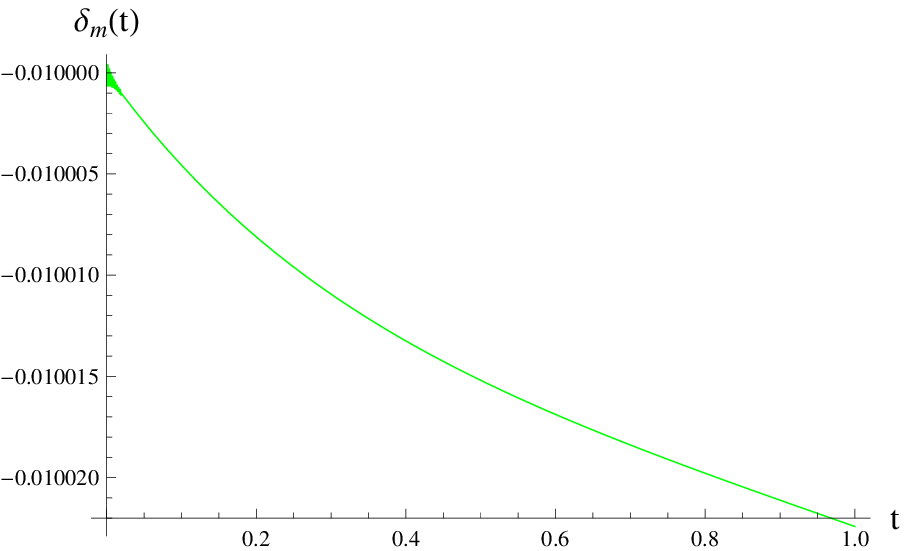,
width=0.5\linewidth}\caption{Evolution of perturbations $\delta(t)$
and $\delta_{m}(t)$ for model (\ref{9d}) with $\omega=0$.}
\end{figure}
\begin{figure}
\epsfig{file=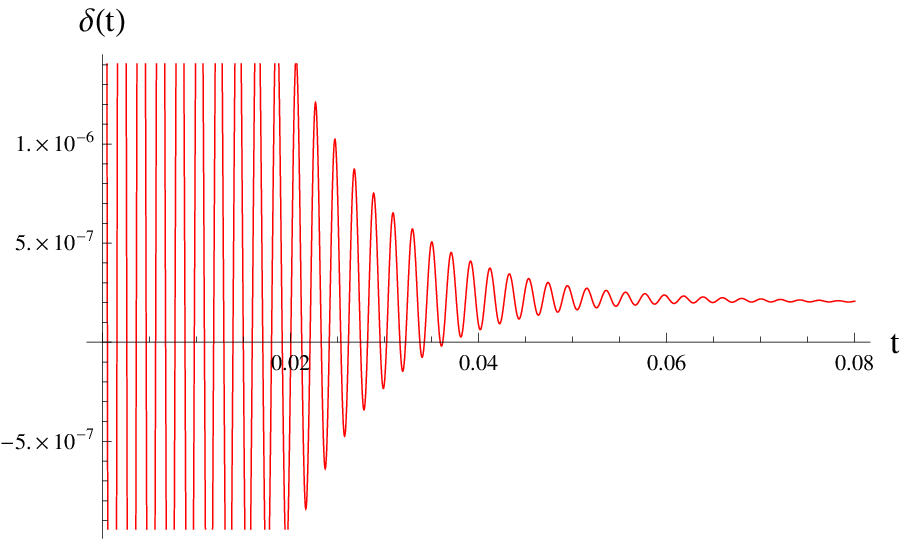, width=0.5\linewidth}\epsfig{file=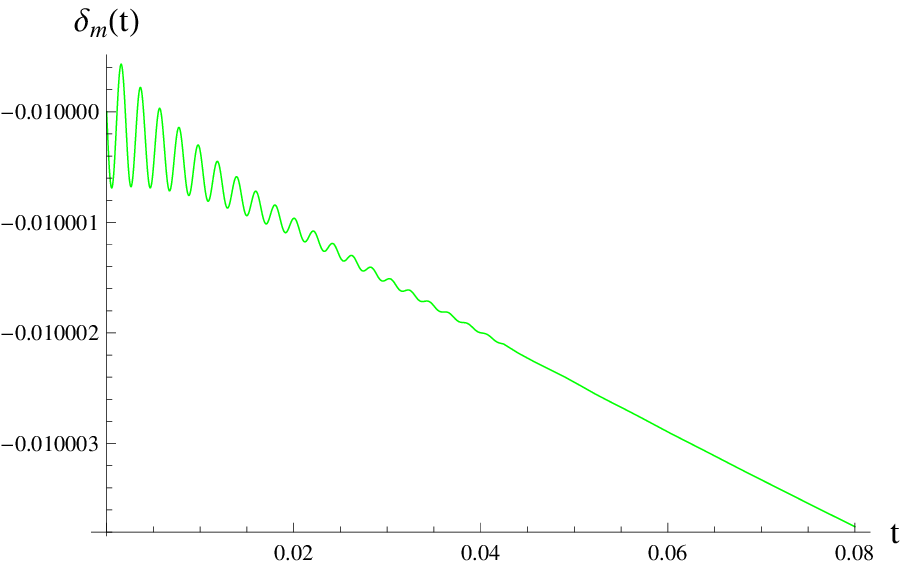,
width=0.5\linewidth}\caption{Evolution of perturbations $\delta(t)$
and $\delta_{m}(t)$ for model (\ref{9d}) with $\omega=0$.}
\end{figure}

\subsection{Stability of Power-law Solutions}

Here we investigate the stability of power-law solutions. These
solutions describe the accelerated as well as decelerated
cosmological evolutionary phases in the background of FRW universe.
We first consider the reconstructed power-law solution (\ref{5p})
and numerically solve Eqs.(\ref{8t}) and (\ref{10t}). For this
model, we choose integration constants
$\tilde{c}_{1}=10,~\tilde{c}_{2}=-0.5$ and $\tilde{c}_{3}=-1000.$
Figure \textbf{5} shows the oscillating behavior of perturbed
parameters $(\delta(t),~\delta_{m}(t))$ for the cosmic accelerated
era with $\omega=-0.5$ and $\lambda=2$. The perturbations around the
power-law solutions decay in future leading to stable results. The
radiation ($\lambda=1/2$ and $\omega=1/3$) as well as matter
($\lambda=2/3$ and $\omega=0$) dominated eras cannot be discussed
for the model (\ref{5p}) because singular as well as complex terms
appear which lead to non-physical case.

Secondly, we consider the model (\ref{6p}) and analyze its behavior
against linear perturbations. Figure \textbf{6} shows the
fluctuating behavior of considered perturbations in the cosmic
accelerated phase with $\omega=-0.5$ and $\lambda=2$. Here, we
choose $\tilde{c}_{1}=0.001,~\tilde{c}_{2}=-0.61$ and
$\tilde{c}_{3}=-1000.$ It is observed that the oscillating behavior
disappears in future while both perturbation parameters will not
decay in late times leading to unstable cosmological solutions. The
considered model cannot explain the cosmological evolution
corresponding to matter and radiation dominated eras like previous
model (\ref{5p}).
\begin{figure}
\epsfig{file=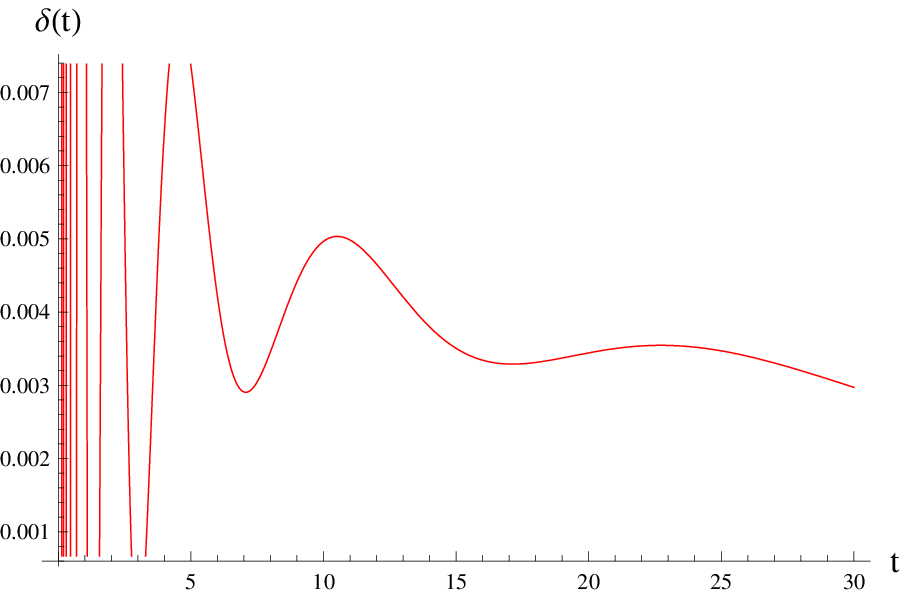, width=0.5\linewidth}\epsfig{file=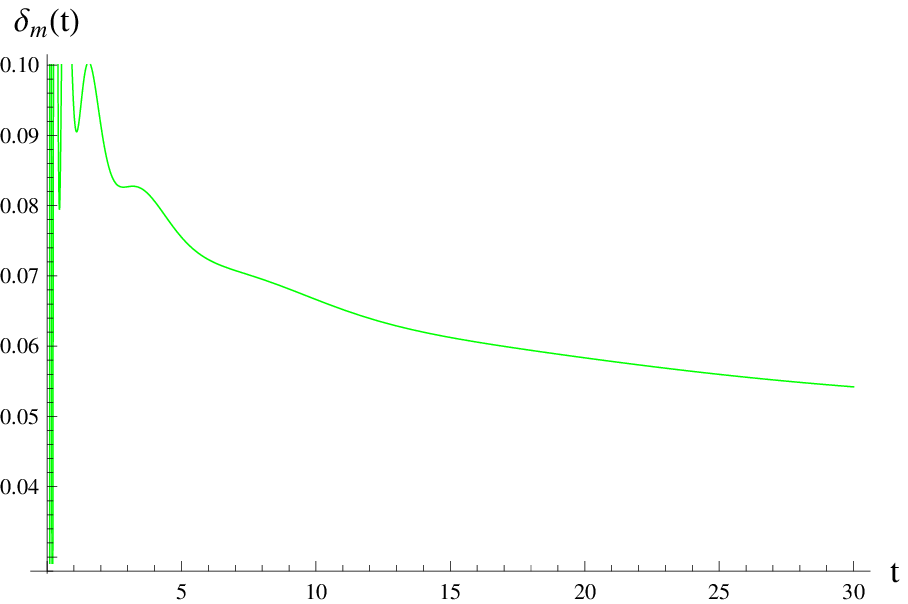,
width=0.5\linewidth}\caption{Evolution of perturbations $\delta(t)$
and $\delta_{m}(t)$ for model (\ref{5p}) with $\omega=-0.5$ and
$\lambda=2$.}
\end{figure}
\begin{figure}
\epsfig{file=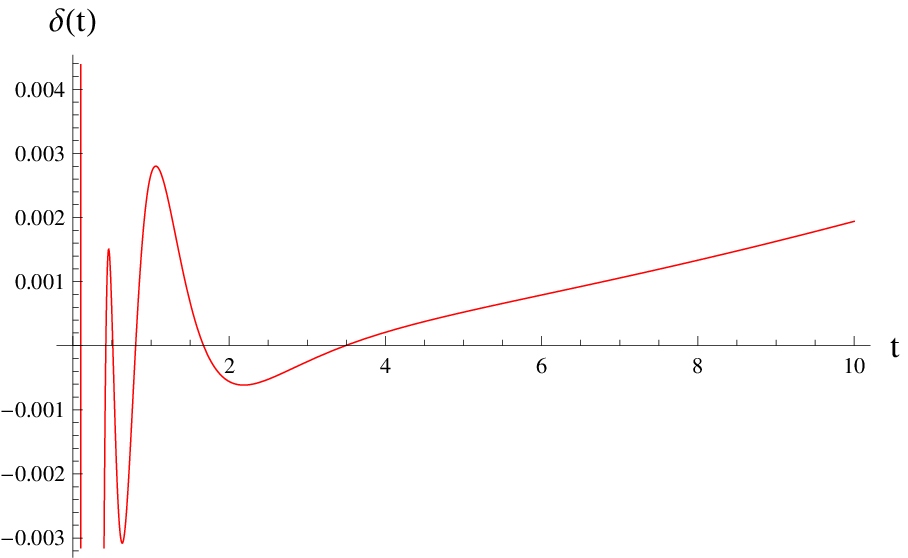, width=0.5\linewidth}\epsfig{file=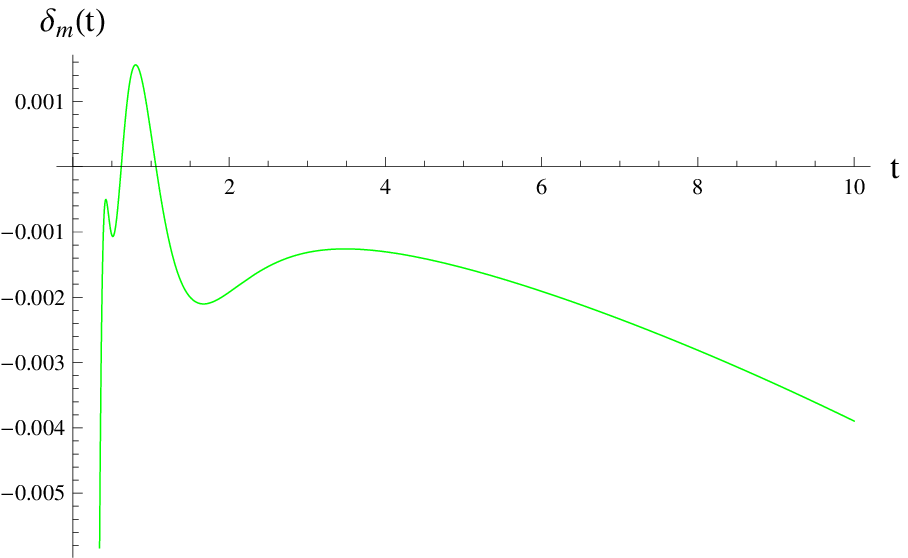,
width=0.5\linewidth}\caption{Evolution of perturbations $\delta(t)$
and $\delta_{m}(t)$ for model (\ref{6p}) with $\omega=-0.5$ and
$\lambda=2$.}
\end{figure}
\begin{figure}
\epsfig{file=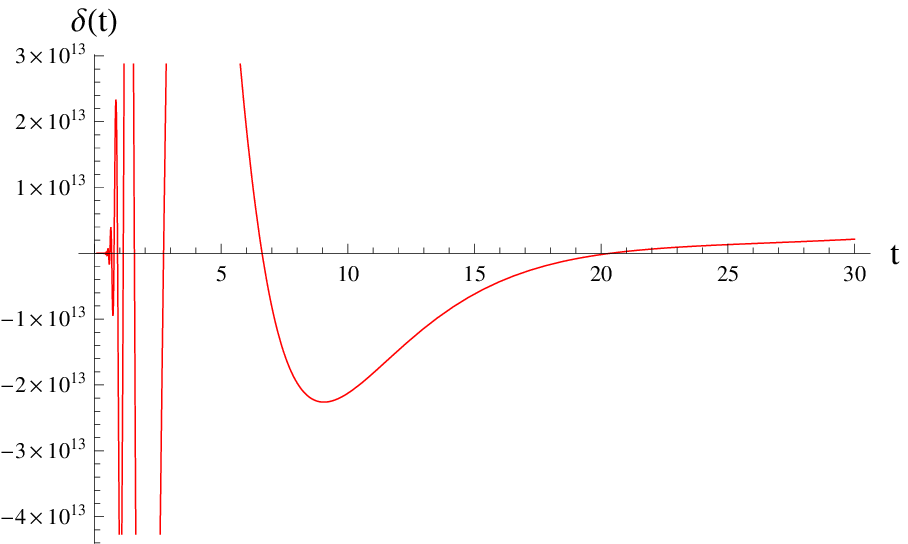, width=0.5\linewidth}\epsfig{file=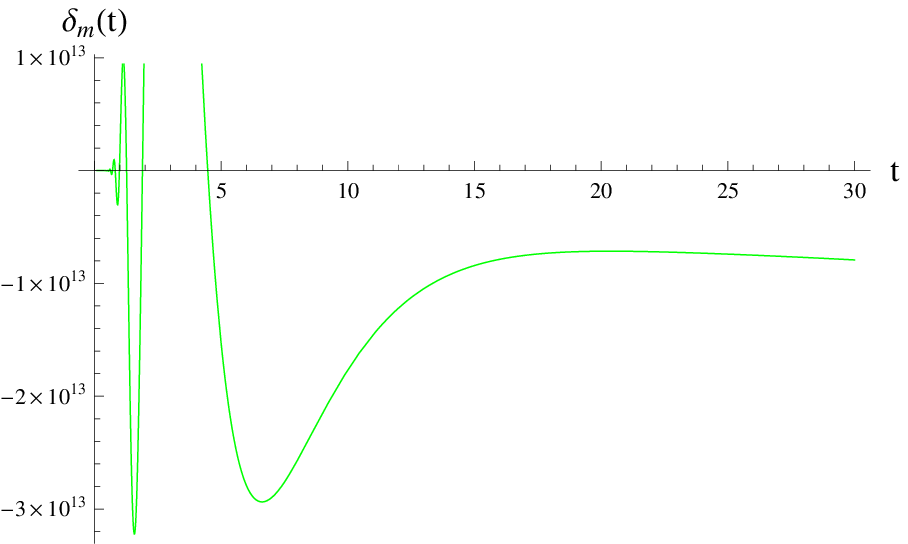,
width=0.5\linewidth}\caption{Evolution of perturbations $\delta(t)$
and $\delta_{m}(t)$ for model (\ref{7p}) with $\omega=-0.5$ and
$\lambda=1.1$.}
\end{figure}

Lastly, we explore the stability of model (\ref{7p}) with
integration constants
$\bar{c}_{1}=-2,~\bar{c}_{2}=-0.6,~\bar{c}_{3}=1000$ and
$\bar{c}_{4}=0.01$. Figure \textbf{7} represents the evolution of
($\delta,~\delta_{m}$) versus time for $\lambda=1.1$ with
$\omega=-0.5$. The left panel shows that the oscillations of
$\delta(t)$ decay in late times while fluctuations of
$\delta_{m}(t)$ remain present in future. Since a complete
perturbation against any cosmological solution includes the matter
perturbations therefore, the solutions are unstable.

\section{Concluding Remarks}

In this paper, we have employed the reconstruction scheme to
$f(\mathcal{G},T)$ gravity in the background of isotropic and
homogeneous universe model to reproduce some important cosmological
models. The basic aspect of this modified gravity is the coupling
between curvature and matter components which yields non-zero
divergence of the energy-momentum tensor. We have imposed additional
constraint to obtain the standard conservation equation which has
been used to explain the cosmic evolution in this gravity.

The de Sitter and power-law solutions have been reconstructed for
general as well as particular cases which are of great interest and
have significant importance in cosmology. We have also reconstructed
the $f(\mathcal{G},T)$ model which can explain cosmic history of the
phantom as well as non-phantom phases of the universe. Similar
reconstruction technique is carried out for $\Lambda$CDM model and
found that this gravity fails to reproduce it for both general as
well as particular $f(\mathcal{G},T)$ forms. The results are
summarized in Table \textbf{1}. In this table, $\checkmark$ and
$\times$ represent that $f(\mathcal{G},T)$ gravity reproduces and
fails to reproduce the corresponding cosmological backgrounds,
respectively.
\newpage
\textbf{Table 1:} Cosmological evolution in $f(\mathcal{G},T)$
gravity.
\begin{table}[bht]
\centering
\begin{small}
\begin{tabular}{|c|c|c|c|}
\hline\textbf{Cosmological Backgrounds}&\textbf{General
$f(\mathcal{G},T)$
Model}&\textbf{Particular $f(\mathcal{G},T)$ Model}\\
\hline de Sitter Universe&$\checkmark$&$\checkmark$\\
\hline Power-law Solutions&$\checkmark$&$\checkmark$\\
\hline Phantom/non-Phantom Eras&$\times$&$\checkmark$\\
\hline
\end{tabular}
\end{small}
\end{table}

On physical grounds, the stability analysis of different forms of
generic function leads to classify the modified theories of gravity.
We have applied the first order perturbations to Hubble parameter
and energy density to analyze the stability of models which
reproduce de Sitter and power-law cosmic history. We have perturbed
the field equation as well as conservation law whose numerical
solutions provide the stable/unstable results.
\begin{itemize}
\item For the de Sitter universe, the evolution of perturbation has been plotted
against time as shown in Figures \textbf{1}-\textbf{4}. These
indicate that models (\ref{6d}) and (\ref{9d}) are stable against
linear perturbations.
\item For the power-law universe, the stability analysis is given in
Figures \textbf{5}-\textbf{7}. It is found that $f(\mathcal{G},T)$
gravity fails to reproduce matter and radiation dominated eras while
stable results are obtained for accelerated phase of the universe
for model (\ref{5p}).
\end{itemize}
We conclude that the cosmological reconstruction and stability
analysis might restrict $f(\mathcal{G},T)$ gravity in the background
of FRW universe. It would be interesting to discuss ghost
instabilities due to the presence of curvature-matter coupling.

\vspace{0.3cm}

\renewcommand{\theequation}{A\arabic{equation}}
\setcounter{equation}{0}
\section*{Appendix A}

The expressions for $\Xi_{i}$'s in Eqs.(\ref{5d}) and (\ref{6d}) are
\begin{eqnarray}\nonumber
\Xi_{1}&=&18c_{1}H_{0}^{4}[8c_{1}H_{0}^{4}\{2(2+59\omega^{2})
-11\omega(5-3\omega^{2})\}-(1-11\omega)(1-\omega^{2})]\\\nonumber
&\times&[1+\omega-36c_{1}H_{0}^{4}(1-3\omega)]^{-2},\\\nonumber
\Xi_{2}&=&18c_{1}H_{0}^{4}[(1-11\omega)(1-\omega^{2})-8c_{1}
H_{0}^{4}\{2(2+59\omega^{2})-11\omega(5-3\omega^{2})\}]\\\nonumber
&\times&[(1+\omega)(1-24c_{1}H_{0}^{4})\{1+\omega-6c_{1}H_{0}^{4}
(5-4\omega-33\omega^{2})\}]^{-1},\\\nonumber\Xi_{3}&=&-[18c_{1}
H_{0}^{4}(1-32c_{1}H_{0}^{4})-3\omega\{1-6c_{1}H_{0}^{4}(3-352c_{1}
H_{0}^{4})\}\\\nonumber&-&2\omega^{2}\{1-9c_{1}H_{0}^{4}
(7-1248c_{1}H_{0}^{4})\}+\omega^{3}\{1-54c_{1}H_{0}^{4}(7-480c_{1}
H_{0}^{4})\}]\\\nonumber&\times&[(1-3\omega)(1-24c_{1}H_{0}^{4})
\{1+\omega-6c_{1}H_{0}^{4}(5-4\omega-33\omega^{2})\}]^{-1}.
\end{eqnarray}
The values for $\gamma_{\hat{j}}$'s in Eq.(\ref{4p}) are
\begin{eqnarray}\nonumber
\gamma_{1}&=&\frac{1}{2}\left[5-\lambda\{1+3\tilde{c}_{2}(1+\omega)\}
\right],\\\nonumber\gamma_{2}&=&\left[\frac{3}{4}\lambda
\tilde{c}_{2}(1+\omega)\{3\tilde{c}_{2}\lambda(1+\omega)+2(\lambda-1)
-8\}+\frac{1}{4}(\lambda-1)(\lambda+7)+4\right.\\\nonumber&+&\left.
8\tilde{c}_{2}(\lambda-1)\left(\frac{1+\omega}{1-3\omega}\right)
\right]^{\frac{1}{2}},\quad\gamma_{3}=-\frac{1}{2}\left(\frac{1-3\omega}
{1+\omega}\right),\quad\gamma_{4}=\frac{2\kappa^2}{\omega-3},\\\nonumber
\gamma_{5}&=&\left(\frac{18\lambda^{3}(1-3\omega)^{\frac{3\lambda
(1+\omega)-2}{3\lambda(1+\omega)}}}{3\lambda(1-3\omega)+4}\right)
\rho_{0}^{\frac{-2}{3\lambda(1+\omega)}},\quad\gamma_{6}=\frac{2}
{3\lambda(1+\omega)}.
\end{eqnarray}
The expressions for $\Delta_{k}$'s and $\Omega_{k}$'s in
Eqs.(\ref{5p}) and (\ref{6p}) are
\begin{eqnarray}\nonumber
\Delta_{1}&=&1-\frac{\gamma_{7}}{\gamma_{8}},\quad\Delta_{2}
=1-\frac{\gamma_{3}^{2}}{\gamma_{8}},\quad\Delta_{3}=\gamma_{4}
\left(1-\frac{1}{\gamma_{8}}\right),\quad\Delta_{4}=\gamma_{5}
\left(1-\frac{\gamma_{6}^{2}}{\gamma_{8}}\right),\\\nonumber
\Omega_{1}&=&1-\frac{\gamma_{8}}{\gamma_{7}},\quad\Omega_{2}
=1-\frac{\gamma_{3}^{2}}{\gamma_{7}},\quad\Omega_{3}=\gamma_{4}
\left(1-\frac{1}{\gamma_{7}}\right),\quad\Omega_{4}=\gamma_{5}
\left(1-\frac{\gamma_{6}^{2}}{\gamma_{7}}\right),
\end{eqnarray}
where
\begin{eqnarray}\nonumber
\gamma_{7}&=&\frac{\tilde{c}_{2}}{6\lambda}\left[6\tilde{c}_{2}
\lambda(1+\omega)^{2}-3\lambda(1+5\omega+2\omega^2)+2(\gamma_{1}
+\gamma_{2})\right],\\\nonumber\gamma_{8}&=&\frac{\tilde{c}_{2}}
{6\lambda}\left[6\tilde{c}_{2}\lambda(1+\omega)^{2}-3\lambda
(1+5\omega+2\omega^2)+2(\gamma_{1}-\gamma_{2})\right].
\end{eqnarray}
The values of $\chi_{h}$'s in Eq.(\ref{8t}) are given as follows
\begin{eqnarray}\nonumber
\chi_{1}&=&288H_{*}^{6}f_{\mathcal{GG}}^{*},\\\nonumber\chi_{2}
&=&288H_{*}^{5}(3H_{*}^{2}+5\dot{H}_{*})f_{\mathcal{GG}}^{*}
+6912H_{*}^{7}(4H_{*}^{2}\dot{H}_{*}+2\dot{H}_{*}^{2}+H_{*}
\ddot{H}_{*})f_{\mathcal{GGG}}^{*}\\\nonumber&+&24(1+\omega
)\rho_{*}H_{*}^{3}f_{\mathcal{G}T}^{*}-864(1+\omega)(1-3\omega)
H_{*}^{7}\rho_{*}f_{\mathcal{GG}T}^{*},\\\nonumber\chi_{3}&=&
-6H_{*}^{2}-24H_{*}^{2}\dot{H}_{*}f_{\mathcal{G}}^{*}-288H_{*}^{4}
(4H_{*}^{4}-23H_{*}^{2}\dot{H}_{*}-11\dot{H}_{*}^{2}-6H_{*}
\ddot{H}_{*})f_{\mathcal{GG}}^{*}\\\nonumber&+&6912H_{*}^{6}
(4H_{*}^{2}+\dot{H}_{*})(4H_{*}^{2}\dot{H}_{*}+2\dot{H}_{*}^{2}
+H_{*}\ddot{H}_{*})f_{\mathcal{GGG}}^{*}+12(1+\omega)\rho_{*}
H_{*}^{2}\\\nonumber&\times&[2(4H_{*}^{2}+\dot{H}_{*})-9(1-3\omega)
H_{*}^{2}]f_{\mathcal{G}T}^{*}-864(1+\omega)(1-3\omega)\rho_{*}
H_{*}^{6}(4H_{*}^{2}\\\nonumber&+&\dot{H}_{*})f_{\mathcal{GG}T}^{*},
\\\nonumber\chi_{4}&=&12(1-3\omega)\rho_{*}H_{*}^{3}f_{\mathcal{G}T}^{*}
,\\\nonumber\chi_{5}&=&\kappa^{2}\rho_{*}-\frac{1}{2}(\omega-3)
\rho_{*}f_{T}^{*}+(1-3\omega)(1+\omega)\rho_{*}^{2}f_{TT}^{*}-12
(1-3\omega)\rho_{*}H_{*}^{2}\\\nonumber&\times&[(4+3\omega)H_{*}^{2}
+\dot{H}_{*}]f_{\mathcal{G}T}^{*}+288(1-3\omega)\rho_{*}H_{*}^{4}
(4H_{*}^{2}\dot{H}_{*}+2\dot{H}_{*}^{2}+H_{*}\ddot{H}_{*})
\\\nonumber&\times&f_{\mathcal{GG}T}^{*}-36(1+\omega)
(1-3\omega)^{2}\rho_{*}^{2}H_{*}^{4}f_{\mathcal{G}TT}^{*}.
\end{eqnarray}
The expressions for $\Upsilon_{h}$'s are
\begin{eqnarray}\nonumber
\Upsilon_{1}&=&3(1+\omega)\rho_{*}H_{*}(\kappa^2+f_{T})-12
(1+\omega)\rho_{*}H_{*}\left[3H_{*}^{2}(1-\omega)(4H_{*}^{2}
+\dot{H}_{*})\right.\\\nonumber&-&\left.2(16H_{*}^{2}\dot{H}_{*}
+6\dot{H}_{*}^{2}+3H_{*}\ddot{H}_{*})\right]f_{\mathcal{G}T}^{*}
-72\rho_{*}^{2}H_{*}^{2}(1-3\omega)(1+\omega)^{2}(4H_{*}^{2}
\\\nonumber&+&\dot{H}_{*})f_{\mathcal{G}TT}^{*}+576\rho_{*}H_{*}^{3}
(1+\omega)(4H_{*}^{2}+\dot{H}_{*})(4H_{*}^{2}\dot{H}_{*}+2
\dot{H}_{*}^{2}\\\nonumber&+&4H_{*}\ddot{H}_{*})f_{\mathcal{GG}T}^{*},
\\\nonumber\Upsilon_{2}&=&-12\rho_{*}H_{*}^{2}(1+\omega)\left[3H_{*}^{2}
(1-\omega)-4(2H_{*}^{2}+3\dot{H}_{*})\right]f_{\mathcal{G}T}^{*}
+576\rho_{*}H_{*}^{2}\\\nonumber&\times&(1+\omega)(4H_{*}^{2}
\dot{H}_{*}+2\dot{H}_{*}^{2}+H_{*}\dot{H}_{*})f_{\mathcal{GG}T}^{*}
-72\rho_{*}^{2}H_{*}^{4}(1-3\omega)\\\nonumber&\times&(1+\omega^{2})
f_{\mathcal{G}TT}^{*},\\\nonumber\Upsilon_{3}&=&24\rho_{*}H_{*}^{3}
(1+\omega)f_{\mathcal{G}T}^{*},\\\nonumber\Upsilon_{4}&=&-\frac{3}{2}
\rho_{*}H_{*}(1+\omega)\left[(1-\omega)f_{T}^{*}+2\rho_{*}^{2}
(1+\omega)(1-3\omega)^{2}f_{TTT}^{*}\right]-\frac{15}{2}
\rho_{*}^{2}H_{*}\\\nonumber&\times&(1-3\omega)(1+\omega)^{2}f_{TT}^{*}
+24\rho_{*}H_{*}(1+\omega)(4H_{*}^{2}\dot{H}_{*}+2\dot{H}_{*}^{2}+H_{*}
\ddot{H}_{*})\\\nonumber&\times&\left[f_{\mathcal{G}T}^{*}+\rho_{*}
(1-3\omega)f_{TT\mathcal{G}}^{*}\right],\\\nonumber\Upsilon_{5}
&=&\rho_{*}\left(\kappa^2+\frac{1}{2}(3-\omega)f_{T}^{*}\right)
+(1+\omega)(1-3\omega)\rho_{*}^{2}f_{TT}^{*}.
\end{eqnarray}
For model (\ref{15}), the coefficients of $(\delta,~\delta_{m})$
have the following expressions
\begin{eqnarray}\nonumber
\hat{\chi}_{1}&=&288H_{*}^{6}F_{\mathcal{GG}}^{*},\\\nonumber
\hat{\chi}_{2}&=&288H_{*}^{5}(3H_{*}^{2}+5\dot{H}_{*})
F_{\mathcal{GG}}^{*}+6912H_{*}^{7}(4H_{*}^{2}\dot{H}_{*}+2
\dot{H}_{*}^{2}+H_{*}\ddot{H}_{*})F_{\mathcal{GGG}}^{*},
\\\nonumber\hat{\chi}_{3}&=&-6H_{*}^{2}-24H_{*}^{2}
\dot{H}_{*}F_{\mathcal{G}}^{*}-288H_{*}^{4}(4H_{*}^{4}-23H_{*}^{2}
\dot{H}_{*}-11\dot{H}_{*}^{2}-6H_{*}\ddot{H}_{*})
\\\nonumber&\times&F_{\mathcal{GG}}^{*}+6912H_{*}^{6}(4H_{*}^{2}
+\dot{H}_{*})(4H_{*}^{2}\dot{H}_{*}+2\dot{H}_{*}^{2}+H_{*}
\ddot{H}_{*})F_{\mathcal{GGG}}^{*},\\\nonumber\hat{\chi}_{5}
&=&\kappa^{2}\rho_{*}-\frac{1}{2}(\omega-3)\rho_{*}\mathcal{F}_{T}^{*}
+(1-3\omega)(1+\omega)\rho_{*}^{2}\mathcal{F}_{TT}^{*},\\\nonumber
\hat{\Upsilon}_{1}&=&3(1+\omega)\rho_{*}H_{*}(\kappa^2+\mathcal{F}_{T}),
\\\nonumber\hat{\Upsilon}_{4}&=&-\frac{3}{2}\rho_{*}
H_{*}(1+\omega)\left[(1-\omega)\mathcal{F}_{T}^{*}+2\rho_{*}^{2}
(1+\omega)(1-3\omega)^{2}\mathcal{F}_{TTT}^{*}\right]-\frac{15}{2}
\rho_{*}^{2}H_{*}\\\nonumber&\times&(1-3\omega)(1+\omega)^{2}
\mathcal{F}_{TT}^{*},\\\nonumber\hat{\Upsilon}_{5}&=&\rho_{*}
\left(\kappa^2+\frac{1}{2}(3-\omega)\mathcal{F}_{T}^{*}\right)
+(1+\omega)(1-3\omega)\rho_{*}^{2}\mathcal{F}_{TT}^{*}.
\end{eqnarray}

\end{document}